\documentclass[11pt]{article}
\usepackage{moriond,epsfig}
\usepackage{cite}

\def\be{\begin{equation}}
\def\ee{\end{equation}}
\def\bea{\begin{eqnarray}}
\def\eea{\end{eqnarray}}

\begin{document}
\vspace*{4cm}
 \title{Heavy flavour and quarkonia production measurement in pp and
Pb--Pb collisions at LHC energies with the ALICE detector}

\author{ Renu Bala, for the ALICE Collaboration }

\address{ University of Jammu, Jammu, India}

\maketitle\abstracts{
ALICE is the dedicated heavy-ion experiment at the LHC. Its main physics
goal is to study the properties of strongly-interacting matter at conditions of
high energy density  and high temperature expected to be reached in central Pb--Pb collisions. Charm and beauty quarks
are well-suited tools to investigate this state of matter since they are produced in initial hard scatterings and are therefore generated early in the system evolution and probe its hottest, densest stage. ALICE recorded pp data at $\rm \sqrt{s}$ = 7 TeV and 2.76 TeV and Pb--Pb data at $\rm \sqrt{s}_{\rm NN}$=2.76 TeV in 2010 and 2011. We  present the latest results on heavy flavour and  J/$\psi$ production at both central and forward rapidity.
}
\section{Introduction}
The main goal of the ALICE \cite{alice,ale} experiment is to study the strongly interacting matter in 
conditions of high density and temperature. In such conditions lattice QCD calculations predict
quark de-confinement and the formation of the so called Quark-Gluon
Plasma (QGP) \cite{qgp}.
Heavy flavour particles are sensitive to the properties of the medium formed in heavy-ion collisions. In particular: open charm and beauty mesons are sensitive to the energy density, through
the mechanism of in-medium energy loss of heavy quarks; quarkonium production suppression (which is expected to give information on the medium  temperature) by colour screening was one of the first proposed signatures for QGP
formation \cite{matsui}; charmonium regeneration due to the recombination of initially uncorrelated c and $\rm \bar{c}$ quarks may occur at LHC energies \cite{pbm,thews}. A detailed description of the physics motivations for heavy flavour measurements in heavy-ion collisions can be found in \cite{ale}. \\

\section{Heavy flavour detection in the ALICE experiment}
The ALICE apparatus \cite{alice} has  excellent capabilities for heavy flavour measurements, for both open heavy flavour hadrons and quarkonia. It is composed of a central barrel and a forward muon arm. In the central region ($|\eta| <$ 0.9), the heavy flavour capability of ALICE
relies in a high granularity tracking system made of the Inner Tracking
System (ITS), the Time Projection Chamber (TPC) and the Transition Radiation
Detector (TRD). The particle identification is performed via dE/dx
measurement in the TPC, via time of flight measurement in the Time Of
Flight detector (TOF). Electrons are identified at low $p_{\rm t}$ ($p_{\rm t} <$ 6 GeV/c) by TPC and TOF, while at  intermediate  and high $p_{\rm t}$ ($p_{\rm t} >$ 2 GeV/c),  the TRD and the Electromagnetic Calorimeter (EMCal) is  used. At forward rapidity, heavy flavour production is measured with the
muon spectrometer (-4 $<\eta<$ -2.5).

The analysis is based on proton-proton collisions at $\rm \sqrt{s}$ = 7 TeV and 2.76 TeV and Pb--Pb collisions at $\rm \sqrt{s_{\rm NN}}$ = 2.76 TeV. The proton-proton data sample consists of Minimum Bias (MB) and muon triggers. The former is defined by the presence of a signal in the Silicon Pixel Detector
(two innermost layers of the ITS) or in either of the two VZERO scintillator arrays, in coincidence with the
beam-beam counters placed at both sides of the interaction point. The muon trigger requires, in addition
to a MB event, that a muon with transverse momentum above about 0.5 GeV/c reaches the muon trigger stations. In Pb--Pb collisions, the MB trigger requires the coincidence of signals in the two arrays of the VZERO and in the SPD. The measurement of the centrality is based on the distribution of signals in the VZERO hodoscopes, modeled with a Glauber calculation \cite{prl}.

\section{Open heavy flavour in pp collisions at $\rm \sqrt{s}$ = 2.76 TeV and 7 TeV}

\subsection{ D mesons via hadronic decays} 

\begin{figure}

\psfig{figure=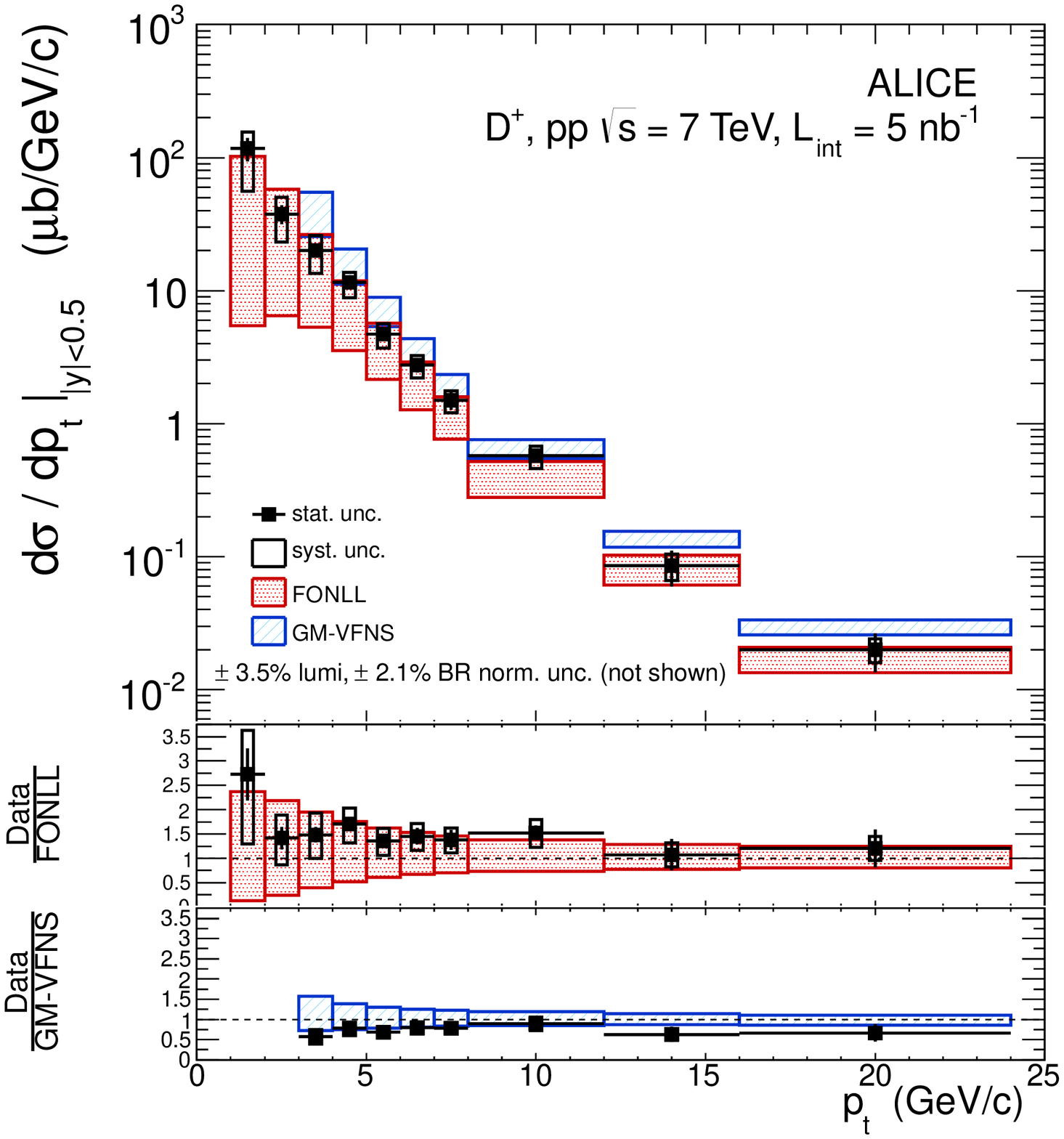,height=2.0in,width=2.0in}
\psfig{figure=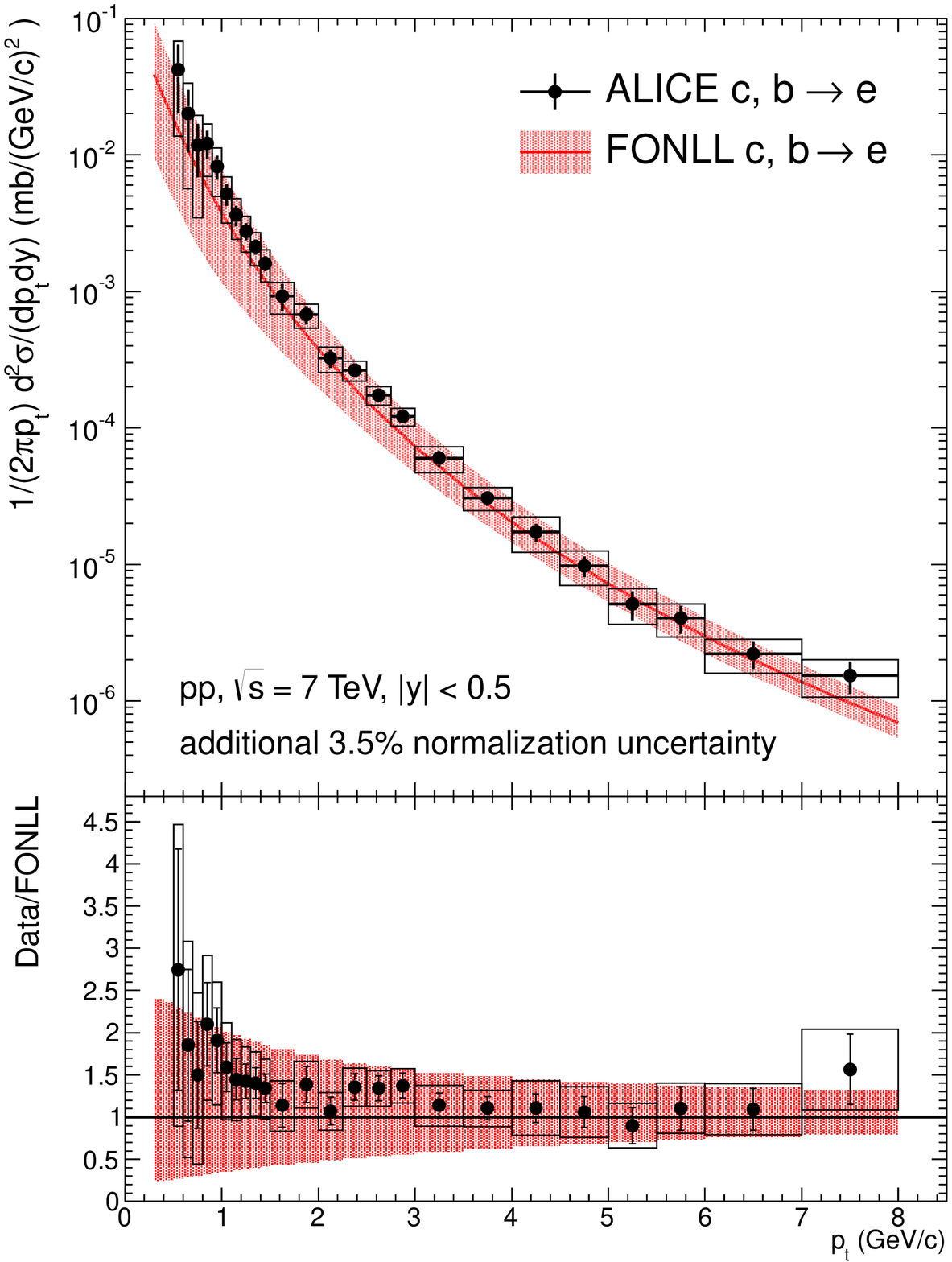,height=2.0in,width=2.0in}
\psfig{figure=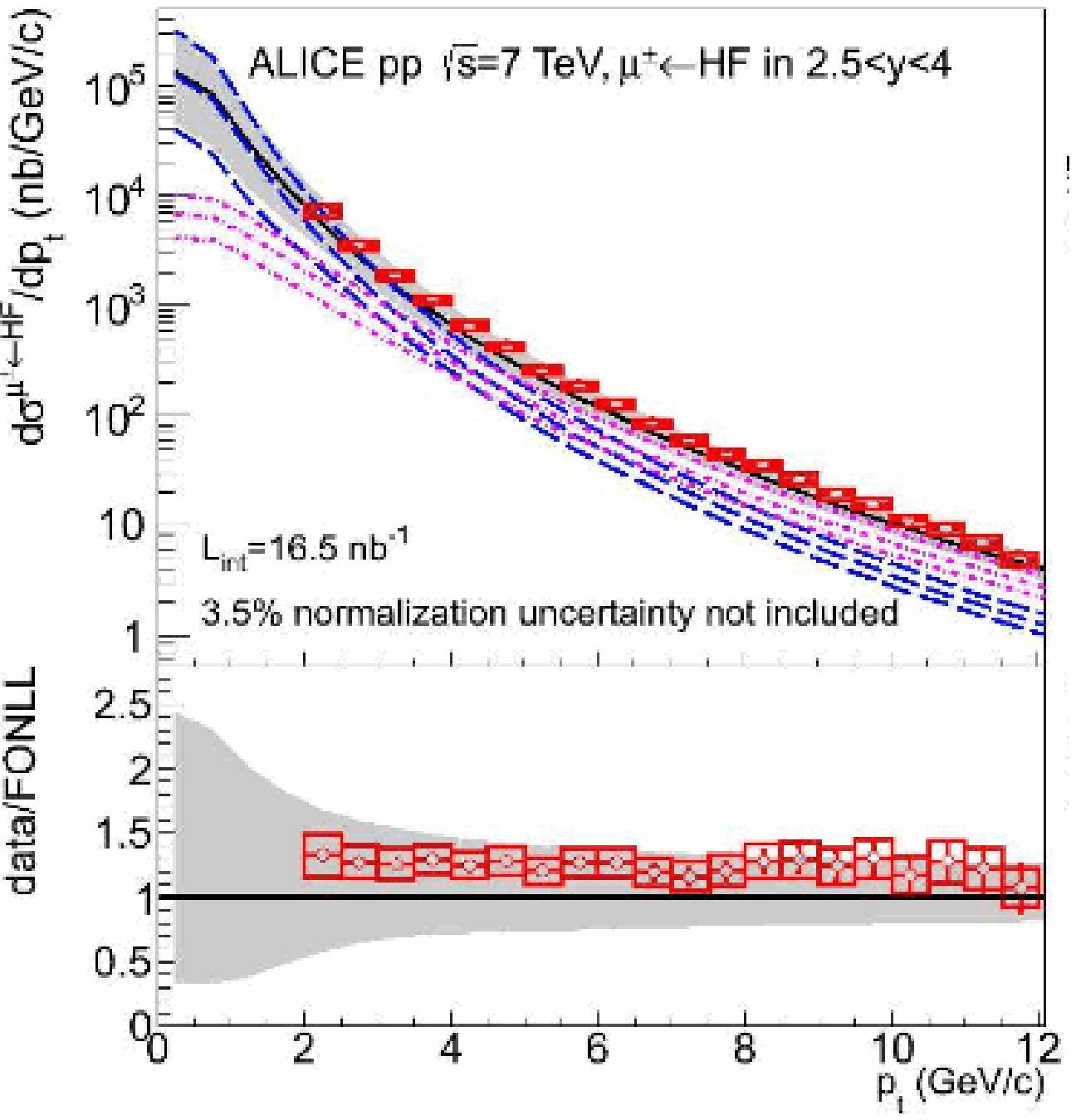,height=2.0in,width=2.0in}

\caption{$p_{\rm t}$ differential cross section for $\rm D^{+}$ mesons in pp collisions at $\rm \sqrt{s}$ = 7 TeV, compared to perturbative QCD predictions from FONLL \cite{fonll,fonll1} and GM-VFNS \cite{gmvfns}  calculations (left) \cite{charmpp}. $p_{\rm t}$ differential cross section of electrons from heavy flavour decays compared to FONLL calculations (middle) \cite{hfe}. $p_{\rm t}$ differential cross section of muons from heavy flavour decays  compared to FONLL calculations (right).} \label{fig:dpp}
\end{figure}

The detection strategy for D mesons at central rapidity is based, for both pp and
Pb--Pb collisions, on the selection of displaced-vertex topologies, i.e. discrimination of
tracks from the secondary vertex from those originating from the primary vertex, large
decay length (normalized to its estimated uncertainty), and good alignment between
the reconstructed D meson momentum and flight-line. The identification of the charged
kaon in the TPC and TOF detectors helps to further reduce the background at low
$ p_{\rm t}$. Figure ~\ref{fig:dpp} (left) shows the $p_{\rm t}$ differential cross section for prompt $\rm D^{+}$ mesons at $ \rm \sqrt{s}$ = 7 TeV. Similar results were obtained for $\rm D^{0}$ and $\rm D^{*+}$ \cite{charmpp}. The feed down from beauty decays is calculated from theory (FONLL) and gives a contribution of 10-15$\%$. The data are well described by pQCD calculations at Fixed-Order Next-to-leading Logarithm level (FONLL) \cite{fonll,fonll1} and GM-VFNS  \cite{gmvfns} predictions. The D mesons cross sections were also measured at $\rm \sqrt{s}$ = 2.76 TeV but with limited statistical precision \cite{charmpp276}. 

\subsection{ Heavy flavour decay electrons and muons}
\begin{figure}
\vskip 2.5cm
\psfig{figure=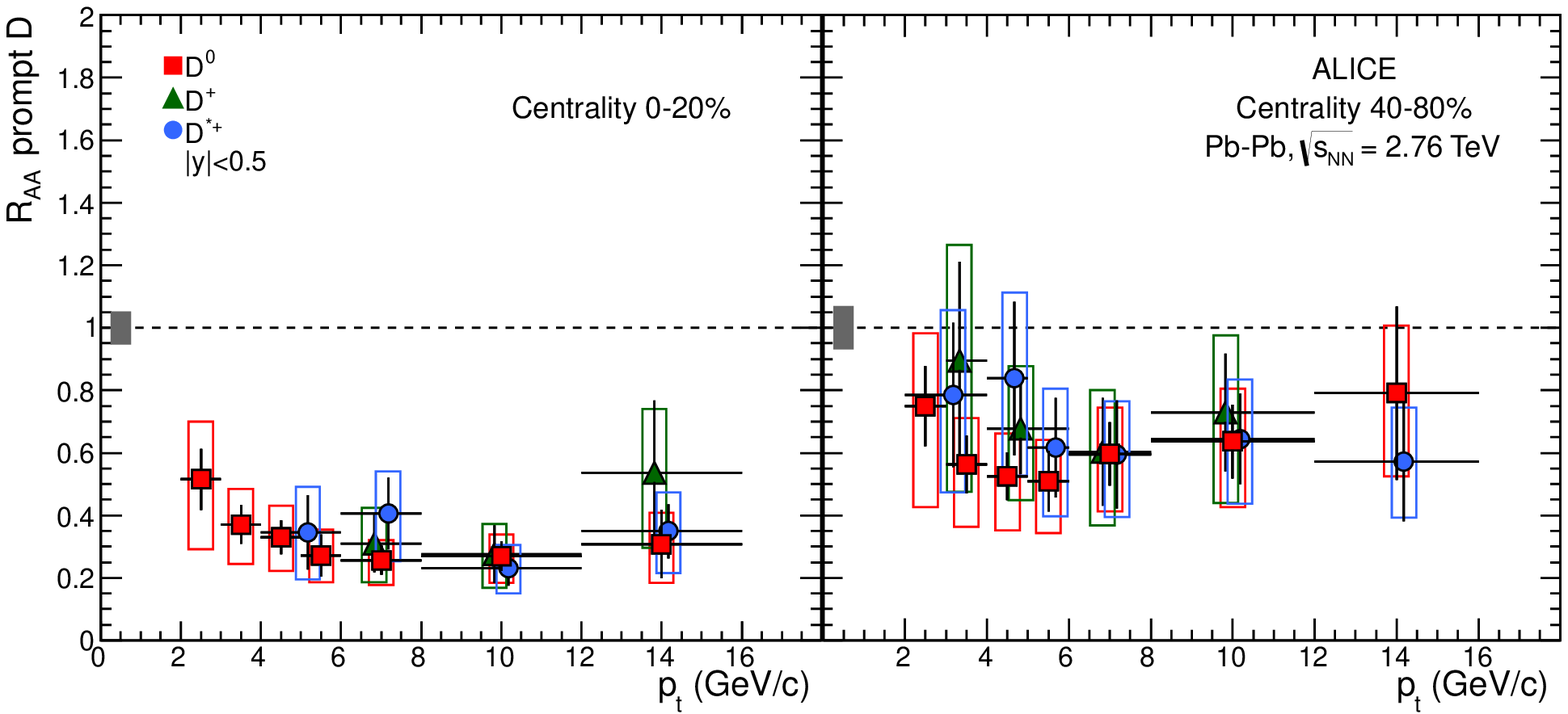,height=2.3in,width=3.5in}
\psfig{figure=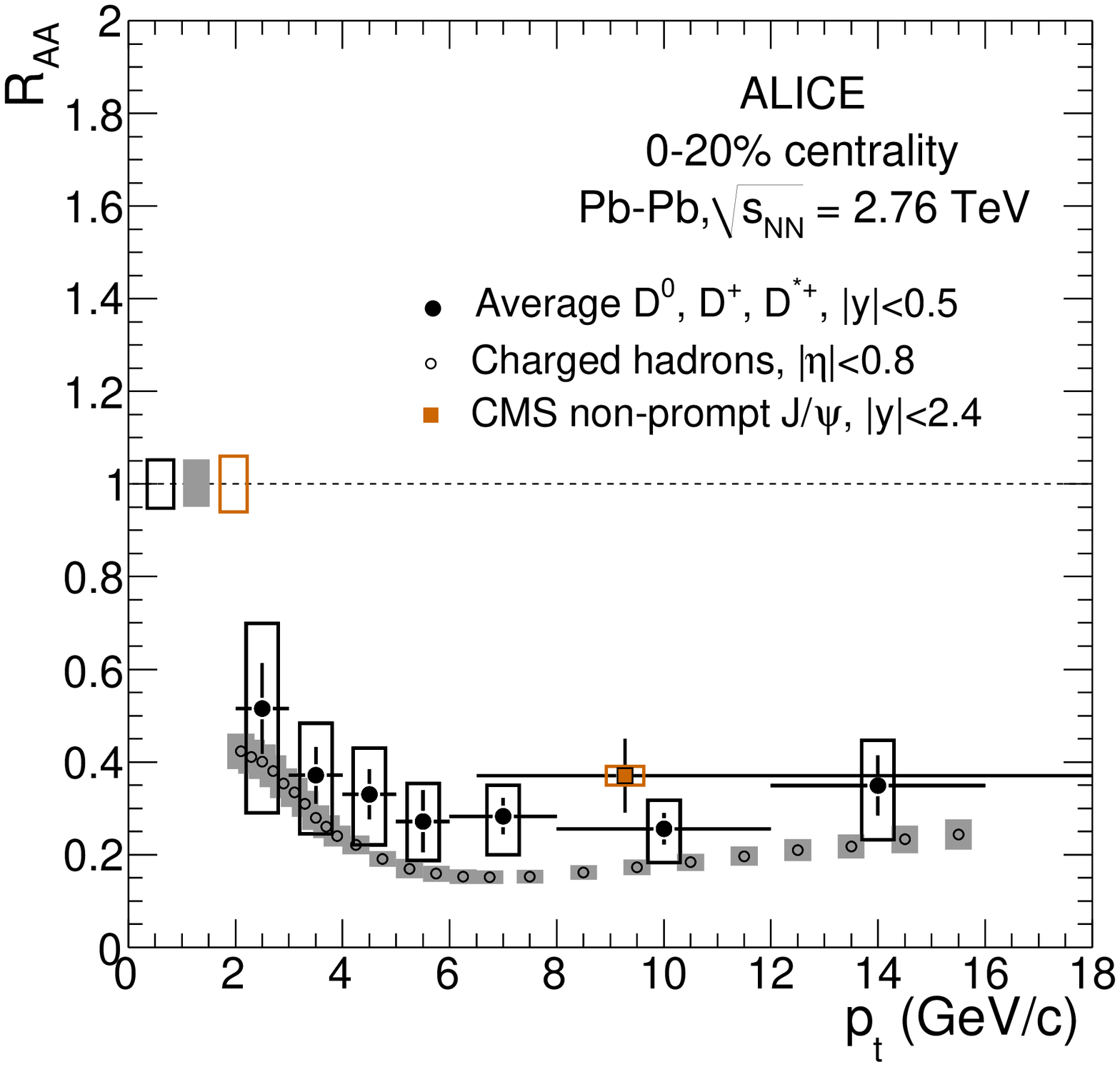,height=2.4in,width=3.0in}
\caption{Prompt $\rm D^{0}$, $\rm D^{+}$ and $\rm D^{*+}$ $\rm R_{AA}$ as a function of $p_{\rm t}$ in 0-20$\%$ and 40-80$\%$ centrality classes (left). Averaged $\rm R_{AA}$ of D-mesons in 0-20$\%$ centrality class compared to the $\rm R_{AA}$ of charged hadrons and non-prompt J/$\psi$ from B-decays (right).} \label{fig:dpbpb}
\end{figure}
At central rapidity, heavy flavour production is measured also using semi-electronic
decays. The key tool for this analysis is the excellent electron PID capability of the ALICE experiment. The TPC dE/dx measurements together with the TOF information allows to identify electrons in
the low and intermediate $p_{\rm t}$ region (up to $\sim$ 4 GeV/c). The analysis includes (at present only for pp data) also the TRD detector to suppress the $\pi$ background. The contribution of electrons from the decay of heavy flavours was extracted from the inclusive electron spectrum by subtracting a cocktail
simulation of the non heavy flavour electron sources. Figure ~\ref{fig:dpp} (middle) shows the heavy flavour electron cross section at $\rm \sqrt{s}$ = 7 TeV compared with FONLL pQCD calculations \cite{hfe}. The data are well described by the theory within uncertainties. 
Heavy flavour production at forward rapidities can be studied in ALICE with single muons.
Muons are measured in the muon spectrometer and identified by requiring that the reconstructed
track matches a tracklet in the trigger system, placed behind an iron wall. This condition
allows to efficiently remove the background contribution of hadrons punching through the frontal
absorber. The main source of background consists of muons from the decay-in-flight of pions and
kaons produced at the interaction point. In pp collisions, such contribution is subtracted through
simulations, while in Pb--Pb collisions, it was estimated by  extrapolating to forward rapidities the $p_{\rm t}$ distributions of pions and kaons  measured at central rapidity. The measured differential production cross sections of muons from heavy flavour decays as a function of $p_{\rm t}$ in the rapidity region 2.5 $<$ y $<$ 4 at $\rm \sqrt{s}$ = 7 TeV is shown in Fig.~\ref{fig:dpp} (right) \cite{hfmu}. Also in this case, the results are well described by FONLL predictions. The muon cross section from heavy flavour decays at $\rm \sqrt{s}$= 2.76 TeV were also measured and details can be found in reference \cite{muPbPb}.

\begin{figure}
\vskip 2.5cm

\psfig{figure=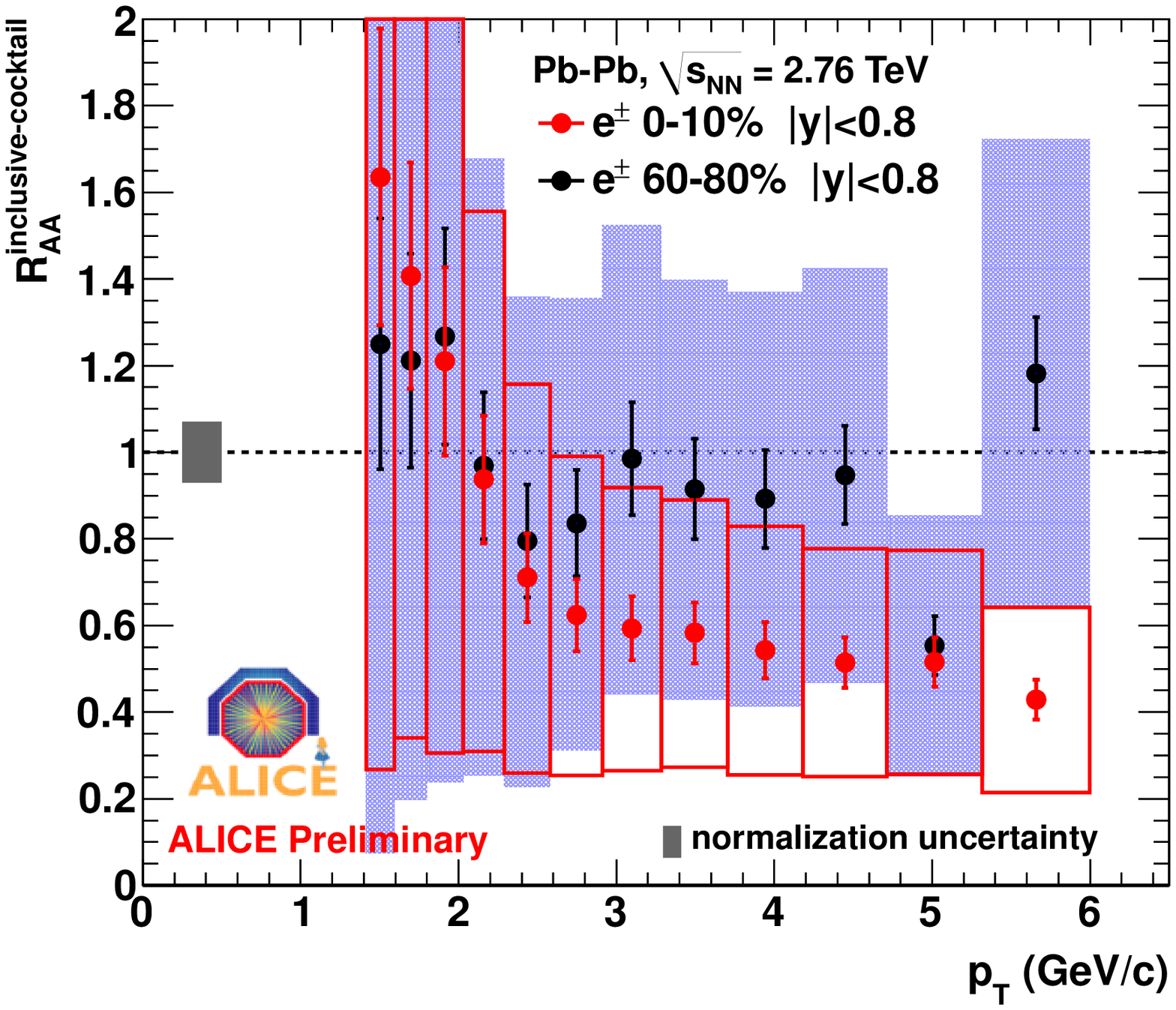,height=2.5in,width=3.0in}
\psfig{figure=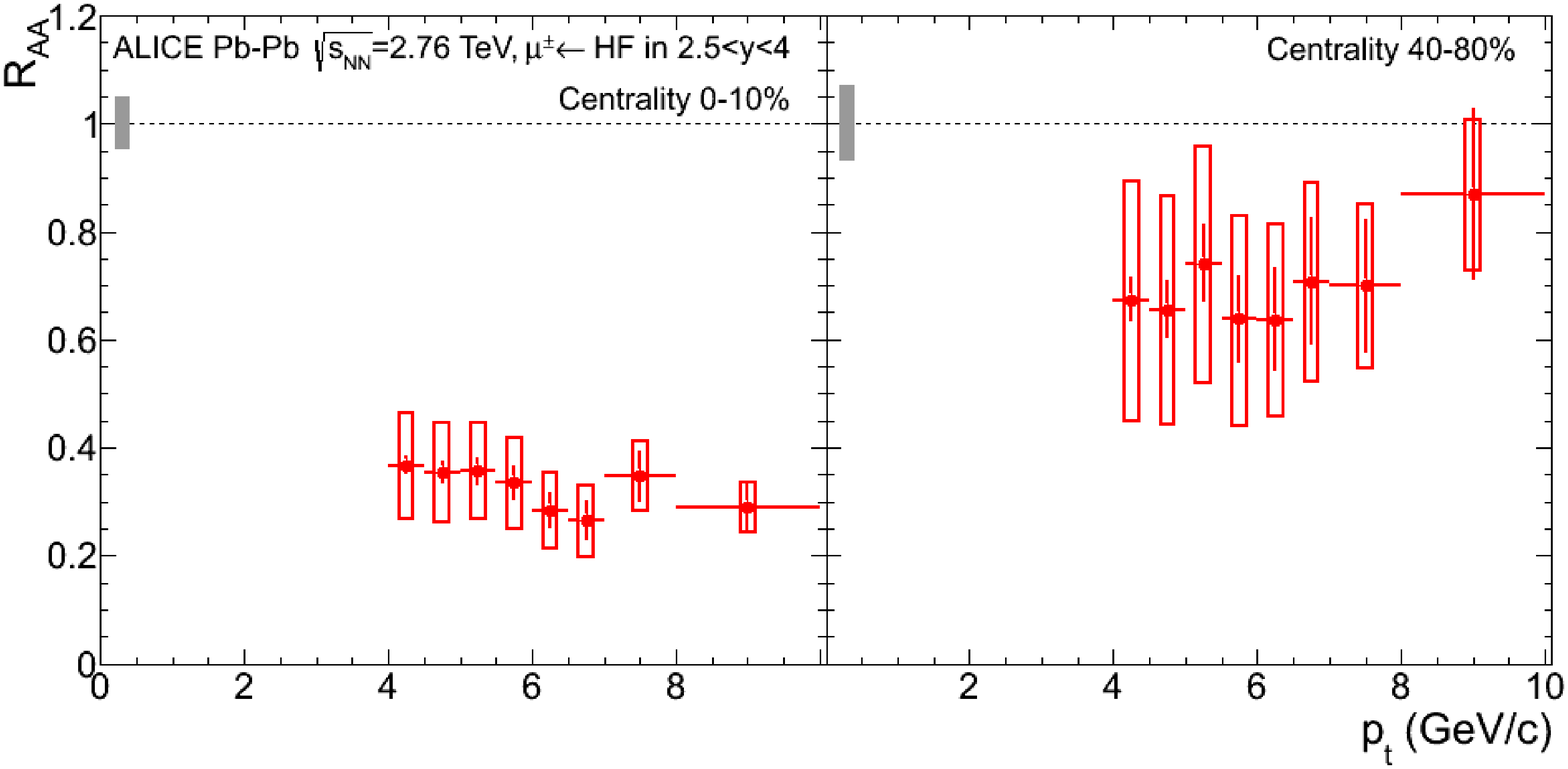,height=2.3in,width=3.7in}
\caption{Comparison of the $\rm R_{AA}$ of background subtracted electrons for central and peripheral Pb--Pb collisions (left). $\rm R_{AA}$ of muons from heavy flavour decays in 2.5 $<$ y $<$ 4 as a function of $p_{\rm t}$ in 0-10$\%$ and 40-80$\%$ centrality classes (right) \cite{muPbPb}.}
\label{fig:muRaa}
\end{figure}
\section{Open heavy flavour in Pb--Pb collisions at $\rm \sqrt{s}$ = 2.76 TeV}

The nuclear modification factor is sensitive to the interaction of hard partons with the medium. It is defined as the ratio of the transverse momentum
spectrum measured in nucleus-nucleus (AA) collisions to the one measured in pp collisions at the
same centre of mass energy, rescaled by the average number of binary nucleon-nucleon collisions
($\rm N_{coll}$) expected in heavy-ion collisions. The ratio can be expressed also in terms of nuclear overlap
integral ($\rm T_{AA}$) estimated within the Glauber-model \cite{prl}.

\begin{center}

$\rm R_{AA}(p_{\rm t}) =\rm \frac{1}{<N_{coll}>}. \rm \frac{\rm dN_{AA}/dp_{\rm t}}{\rm dN_{pp}/dp_{\rm t}} =\rm \frac{1}{<T_{AA}>}
. \frac{\rm dN_{AA}/dp_{\rm t}}{\rm d\sigma_{pp}/dp_{\rm t}}$
\end{center}

 For D mesons and electrons from  heavy flavour decays, the pp reference is scaled from 7 to 2.76 TeV using pQCD calculations (FONLL) \cite{ref}. The scaled results for D mesons were cross checked with the available measurement in pp collisions at $\rm \sqrt{s}$ = 2.76 TeV \cite{charmpp276} (this sample was not used in $\rm R_{AA}$ due to the limited statistics). The systematic uncertainties were obtained by taking into account the full theoretical uncertainties, and assuming no dependence of the quark mass and scales with $\rm \sqrt{s}$. \\
 For the muon $\rm R_{AA}$, the pp reference was obtained from the analysis of muon triggered events collected during a pp run at $\rm \sqrt{s}$ = 2.76 TeV
\subsection{D mesons via hadronic decays}

The transverse momentum dependence of the nuclear modification factor for D-mesons is shown in  
Fig.~\ref{fig:dpbpb} (left) for the central (0-20$\%$ centrality class) and semi-peripheral (40-80$\%$) events. $\rm R_{AA}$ for the three species agree with each other in both centrality classes.
 A clear increase in the $\rm R_{AA}$ is visible for more peripheral collisions. A comparison
of the averaged D meson $\rm R_{AA}$ with the charged hadrons $\rm R_{AA}$ was carried out and is shown in  Fig.~\ref{fig:dpbpb} (right).
Since at high $p_{\rm t}$ ($p_{\rm t} >$ 5 GeV/c) it is shown that the charged hadron $\rm R_{AA}$
coincides with that for charged pions \cite{pion}, the comparison would allow to test
the prediction about the colour charge and mass dependence of energy loss,
according to which heavy quarks would lose less energy than gluons, translating
into $\rm R_{AA}^{D}  > R_{AA}^{charged}$ . The results show comparable suppression for heavy
and light hadrons $\rm R_{AA}$ especially at $p_{\rm t} >$ 5 GeV/c. Nevertheless, there are some
indications that $\rm R_{AA}^{D}$ may be higher than $\rm R_{AA}^{charged}$ at
low $p_{\rm t}$ (up to  $\sim$ 30$\%$  at 3 GeV/c). Also in the same figure, $\rm R_{AA}$ measured by the CMS Collaboration for non-prompt J/$\psi$ mesons (from B decays) with $p_{\rm t} >$ 6.5 GeV/c \cite{cms} is shown.  The non-prompt J/$\psi$ suppression is clearly weaker than that of charged hadrons, while the comparison with D mesons  require more differential and precise
measurements of the transverse momentum dependence. Comparisons with different theoretical models have also been carried out.
Several models describe reasonably well both the charm $\rm R_{AA}$ and the
light  flavour $\rm R_{AA}$. For more details on the analysis, see reference
\cite{charmpbpb}.

\subsection{Single muons and electrons}
Figure ~\ref{fig:muRaa} (left) shows the nuclear modification factor of electrons from heavy flavour decays  for the centrality ranges 0-10$\%$ (central events) and 60-80$\%$ (peripheral events).  In peripheral
Pb-Pb collisions, $\rm R_{AA}$ is compatible with one, whereas a suppression by
a factor 1.2-5 is observed for the most central collisions in the $p_{\rm t}$ region
3.5-6 GeV/c, where charm and beauty decays dominates. 
Figure ~\ref{fig:muRaa} (right) presents the $\rm R_{AA}$ of muons from heavy flavour decays in 2.5 $<$ y $<$ 4, as a function of $p_{\rm t}$ in central (0-10$\%$) and peripheral (40-80$\%$) collisions \cite{muPbPb}. A larger suppression is observed in central collisions than in peripheral collisions with no significant dependence on $p_{\rm t}$ within uncertainties. These heavy flavour decay lepton measurements indicates a 
strong coupling of heavy quarks to the medium created in heavy-ion
collisions. \\

\begin{figure}
\vskip 2.5cm
\psfig{figure=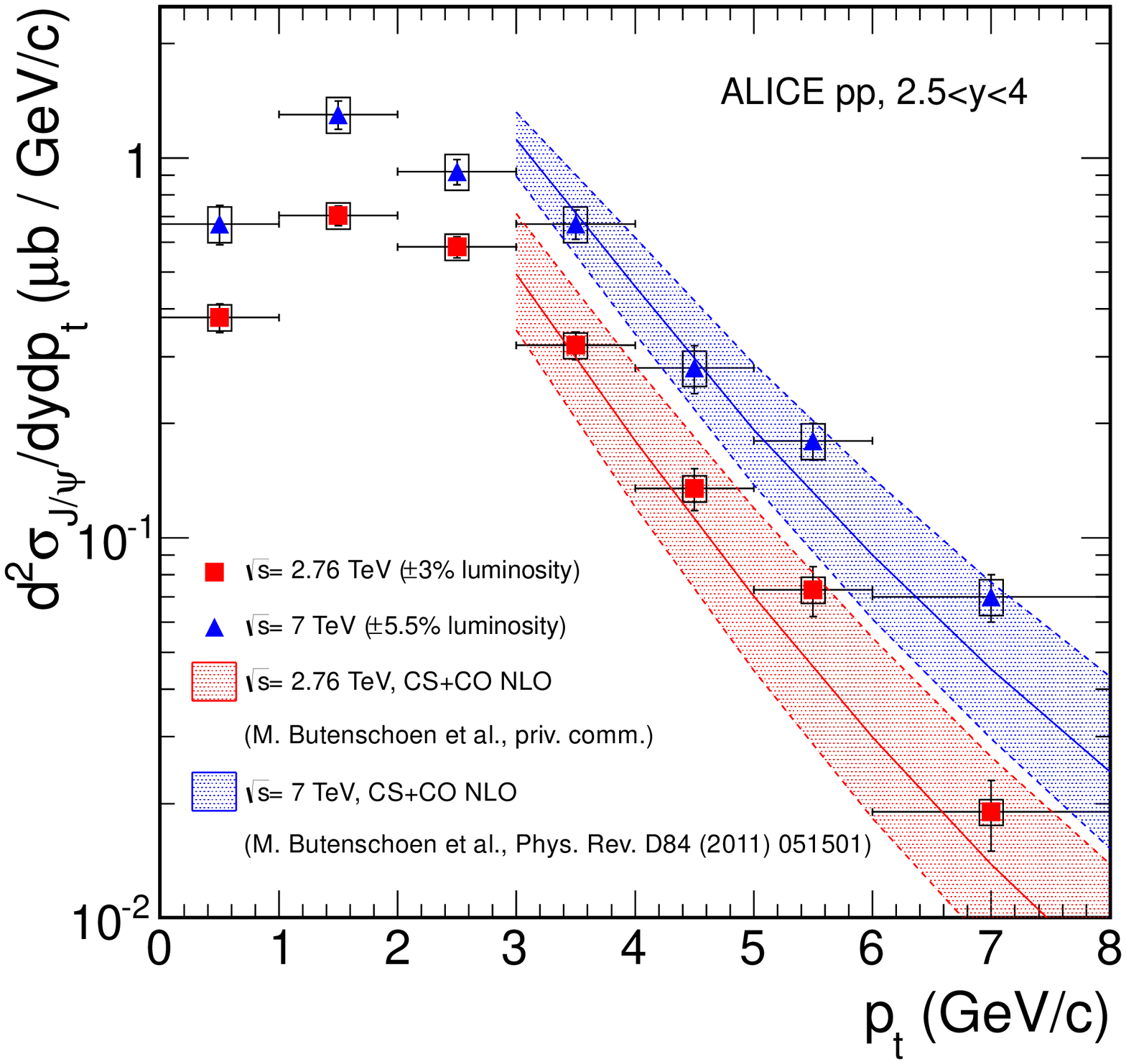,height=2.0in,width=2.0in}
\psfig{figure=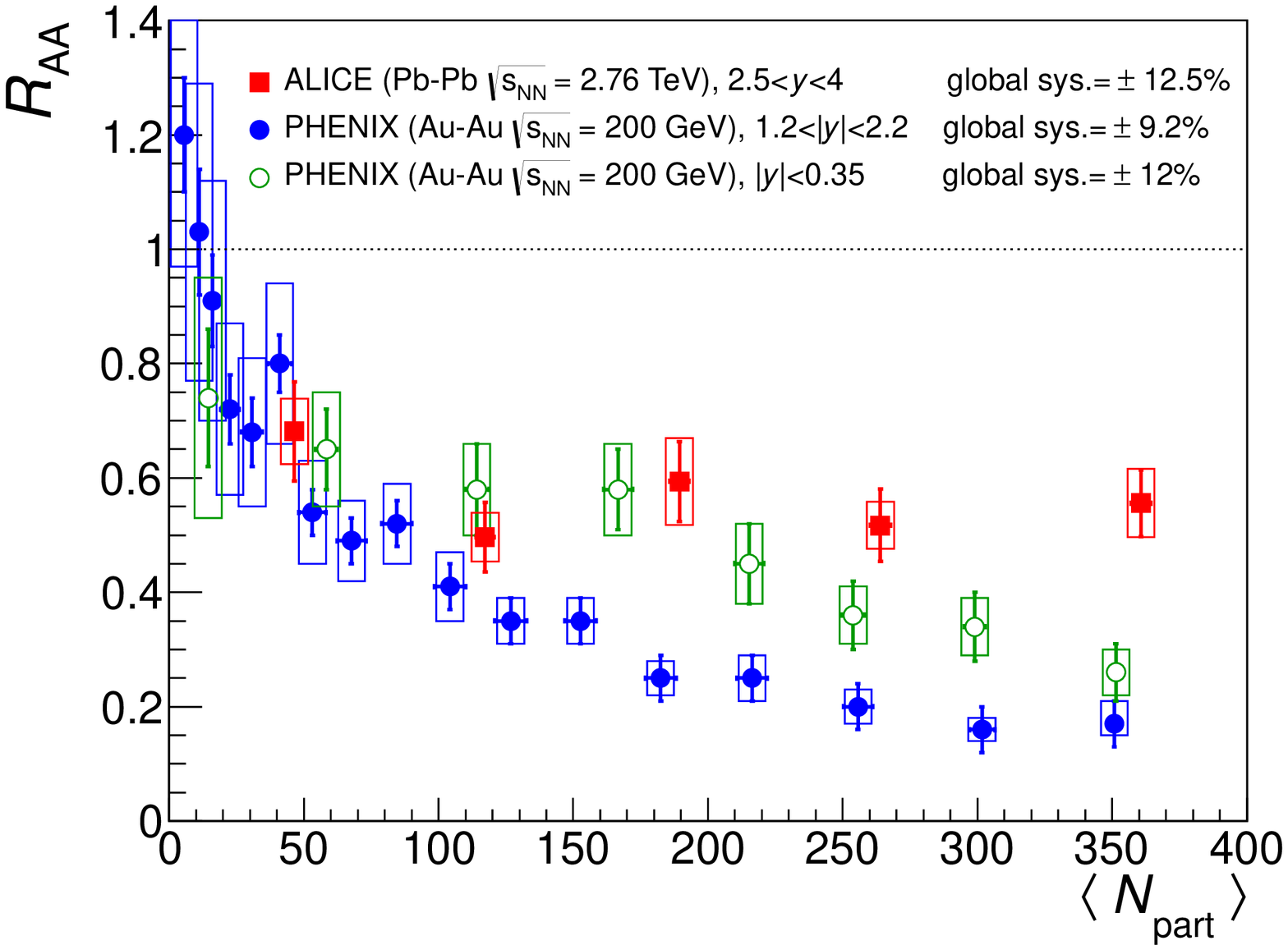,height=2.0in,width=2.0in}
\psfig{figure=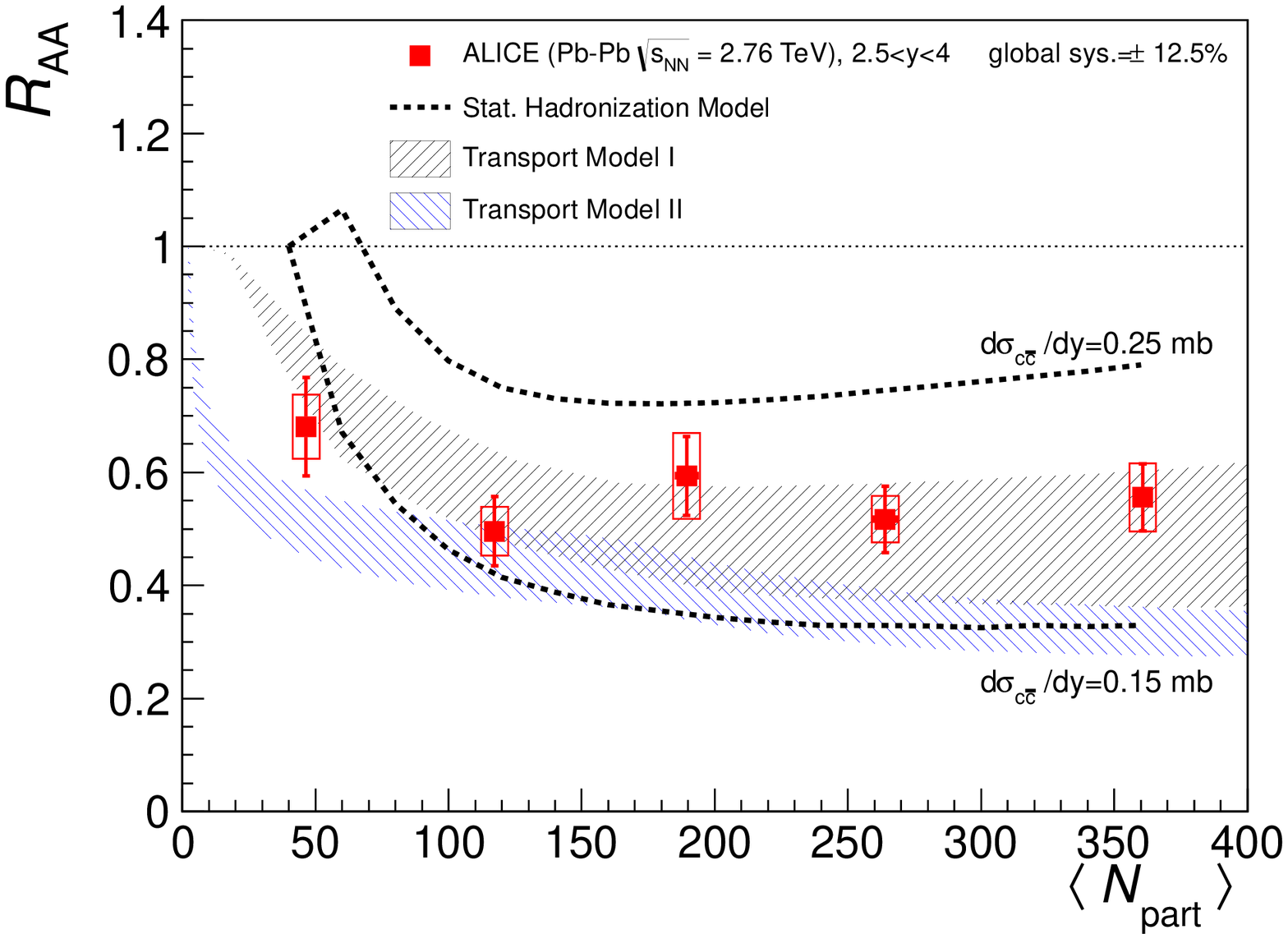,height=2.0in,width=2.0in}

\caption{$p_{\rm t}$ differential cross section for inclusive J/$\psi$ compared to NRQCD calculation (left) \cite{jpsi276}. Inclusive J/$\psi$ $\rm R_{AA}$ as a function of the average number of participating nucleons  measured in Pb--Pb collisions at $\rm \sqrt{s_{\rm NN}}$ = 2.76 TeV compared to PHENIX results in Au--Au collisions at $\rm \sqrt{s_{\rm NN}}$ = 200 GeV (middle) \cite{jpsipbpb}. Inclusive J/$\psi$ $\rm R_{AA}$ compared to the predictions by Statistical Hadronization Model \cite{Hadr}, Transport Model I \cite{Tra1} and II \cite{Tra2} (right).}
\label{fig:jpsipp}
\end{figure}

\section{Quarkonia production measurement in pp and Pb--Pb collisions}

The inclusive J/$\psi$ measurements was performed at $\rm \sqrt{s}$ = 2.76 TeV and 7 TeV. The integrated
and differential cross sections were evaluated down to $p_{\rm t}$ = 0 in two rapidity ranges, $|y| <$ 0.9 and
 2.5 $<$ y $<$ 4, in the dielectron and dimuon decay channel, respectively. The measurement at $\rm \sqrt{s}$
= 2.76 TeV  provides a crucial reference for the study
of hot nuclear matter effects on J/$\psi$
production. Figure ~\ref{fig:jpsipp} (left) shows  the differential cross section, averaged over the interval 2.5 $<$ y $<$ 4 in the transverse momentum range $\rm 0 < p_{\rm t} < 8$ for the two measured energies \cite{jpsi276,jpsi7}. The results are compared with the predictions of a NRQCD calculation \cite{nrqcd} that is in agreement with the data. \\
 The inclusive J/$\psi$ $\rm R_{AA}$ is shown in Fig.~\ref{fig:jpsipp} (middle), for $p_{\rm t}>0$ and 2.5 $<$ y $<$ 4, as a function of the average number of nucleons participating to the collision.  The comparison with the results from PHENIX in Au--Au collisions at $\rm \sqrt{s_{\rm NN}}$ = 0.2 TeV \cite{Phenix1,Phenix2}, which are also shown in the figure, indicates a smaller suppression at LHC than at RHIC in central collisions. Models based on statistical hadronization, or including J/$\psi$ regeneration from charm quarks in the QGP phase can describe the data  as shown in Fig.~\ref{fig:jpsipp} (right) and as described in  reference \cite{jpsipbpb}.

\section{Conclusions}

The latest ALICE results on open heavy flavour and quarkonium measurements in pp and
Pb--Pb were presented. In pp collisions, the open heavy flavour production measurements are well described by NLO pQCD calculations. The J/$\psi$ cross sections were measured in a wide rapidity range down to $p_{\rm t}$ = 0. The $\rm R_{AA}$ measurement shows a large suppression for D
mesons in the centrality 0-20$\%$ as well as for electrons and muons from heavy flavour decays. The D mesons suppression  at $p_{\rm t} >$ 5 GeV/c is compatible with that of charged hadrons. Below 5 GeV/c, there is hint of possible hierarchy in the values of $\rm R_{AA}$ i.e. $\rm R_{AA}^{D} > R_{AA}^{charged}$. The higher statistics from 2011 Pb--Pb run should allow for a firm conclusion. In addition, the comparison data from p--Pb collisions should allow to disentangle initial-state nuclear effects, which could be different for light and heavy flavours.  A significant suppression is also observed in the inclusive J/$\psi$ production in Pb--Pb collisions. The J/$\psi$ $\rm R_{AA}$ is larger than the one measured  at RHIC in central collisions, which could be an indication of (re)generation of J/$\psi$ in the QGP. A better knowledge of the cold nuclear matter effects, to be studied by means of  p--Pb collisions, will be required to constrain suppression/regeneration models.

\end{document}